\documentclass{article}
\usepackage{spconf,amsmath,graphicx,subcaption,flushend}
\usepackage[svgnames]{xcolor}
\usepackage{hyperref}
\hypersetup{
  colorlinks   = true, 
  urlcolor     = DarkBlue, 
  linkcolor    = black, 
  citecolor   = black 
}

\title{Music demixing with the sliCQ transform}
\name{Sevag Hanssian}
\address{McGill University, Montr{\'e}al, Canada}
\begin{document}
\ninept
\maketitle
\begin{abstract}
	Music source separation is the task of extracting an estimate of one or more isolated sources or instruments (for example, drums or vocals) from musical audio. The task of music demixing or unmixing considers the case where the musical audio is separated into an estimate of all of its constituent sources that can be summed back to the original mixture. The Music Demixing Challenge\footnote{\url{https://www.aicrowd.com/challenges/music-demixing-challenge-ismir-2021}} \cite{mdx21} was created to inspire new demixing research. Open-Unmix (UMX) \cite{umx}, and the improved variant CrossNet-Open-Unmix (X-UMX) \cite{xumx}, were included in the challenge as the baselines. Both models use the Short-Time Fourier Transform (STFT) as the representation of music signals.

	The time-frequency uncertainty principle states that the STFT of a signal cannot have maximal resolution in both time and frequency \cite{gabor1946}. The tradeoff in time-frequency resolution can significantly affect music demixing results \cite{tftradeoff1}. Our proposed adaptation of UMX replaced the STFT with the sliCQT \cite{slicq}, a time-frequency transform with varying time-frequency resolution. Unfortunately, our model xumx-sliCQ\footnote{\url{https://github.com/sevagh/xumx-sliCQ}} achieved lower demixing scores than UMX.

\end{abstract}
\section{Introduction}
\label{sec:intro}

The STFT is computed by applying the Discrete Fourier Transform on fixed-size windows of the input signal. From both auditory and musical motivations, variable-size windows are preferred, with long windows in low-frequency regions to capture detailed harmonic information with a high frequency resolution, and short windows in high-frequency regions to capture transients with a high time resolution \cite{doerflerphd}. The sliCQ Transform (sliCQT) \cite{slicq} is a realtime variant of the Nonstationary Gabor Transform (NSGT) \cite{balazs}. These are time-frequency transforms with complex Fourier coefficients and perfect inverses that use varying windows to achieve nonlinear time or frequency resolution. An example application of the NSGT/sliCQT is an invertible Constant-Q Transform (CQT) \cite{jbrown}.

\section{Methodology}
\label{sec:method}

In music demixing, the oracle estimator represents the upper limit of performance using ground truth signals. In UMX, the phase of the STFT is discarded and the estimated magnitude STFT of the target is combined with the phase of the mix for the first estimate of the waveform. This is sometimes referred to as the ``noisy phase'' \cite{noisyphase1}, described by equation \eqref{eq:noisyphaseoracle}.
\begin{equation}\label{eq:noisyphaseoracle}
\hat{X}_{\text{target}} = |X_{\text{target}}| \cdot \angle{X_{\text{mix}}}
\end{equation}

\begin{figure*}[h]
	\includegraphics[width=\textwidth]{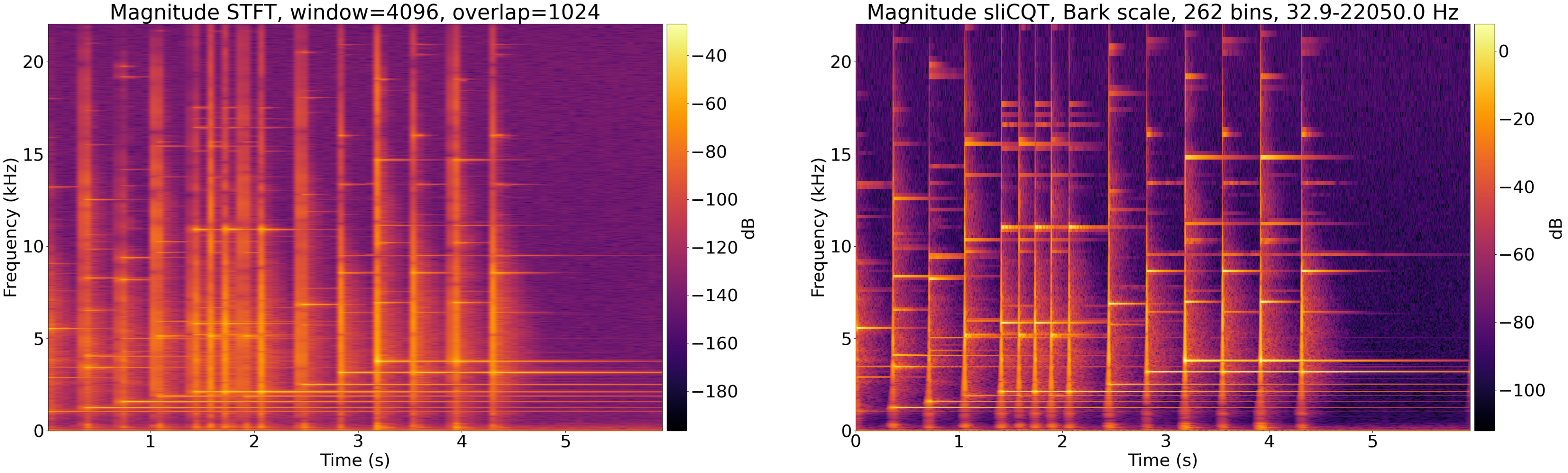}
	\caption{STFT and sliCQT spectrograms of the musical glockenspiel signal}
	\label{fig:spectrograms}
\end{figure*}

The sliCQT parameters were chosen randomly in a 60-iteration search for the largest median SDR across the four targets (vocals, drums, bass, other) from the noisy-phase waveforms of the MUSDB18-HQ \cite{musdb18hq} validation set. The sliCQT parameters of 262 frequency bins on the Bark scale between 32.9--22050 Hz achieved 7.42 dB in the noisy phase oracle, beating the 6.23 dB of the STFT with the UMX window and overlap of 4096 and 1024 samples respectively. STFT and sliCQT spectrograms of a glockenspiel signal\footnote{\url{https://github.com/ltfat/ltfat/blob/master/signals/gspi.wav}} are shown in Figure \ref{fig:spectrograms}.

\begin{figure*}[h]
	\includegraphics[width=\textwidth]{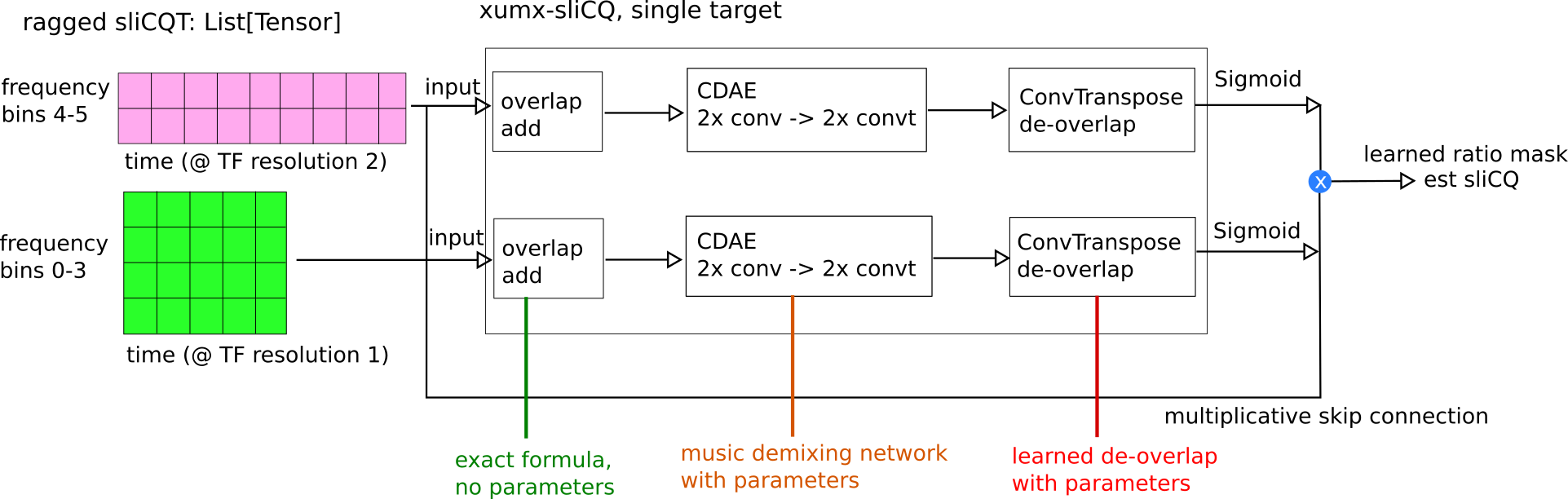}
	\caption{Convolutional denoising autoencoders (CDAE) applied to the ragged sliCQT}
	\label{fig:ragged}
\end{figure*}

The STFT outputs a single time-frequency matrix where all of the frequency bins are spaced uniformly apart and have the same time resolution. The sliCQT groups frequency bins, which may be nonuniformly spaced, in a ragged list of time-frequency matrices, where each matrix contains frequency bins that share the same time resolution. In xumx-sliCQ, a Convolutional Denoising Autoencoder (CDAE) architecture (adapted from STFT-based music source separation models \cite{plumbley1, plumbley2}) was applied separately to each time-frequency matrix, shown in Figure \ref{fig:ragged}. Note how the sliCQT must be overlap-added before being used as an input to the network; the de-overlap was learned with a final transposed convolutional layer after the CDAE layers.

\section{Results}
\label{sec:results}

\begin{figure*}[h]
	\centering
	\subfloat[UMX, single target]{\includegraphics[width=0.75\textwidth]{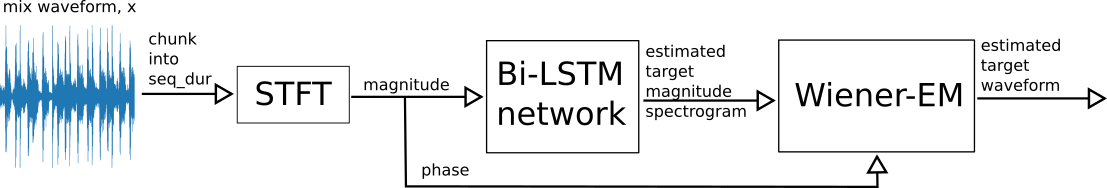}}\\
	\vspace{1em}
	\subfloat[xumx-sliCQ, single target]{\includegraphics[width=0.75\textwidth]{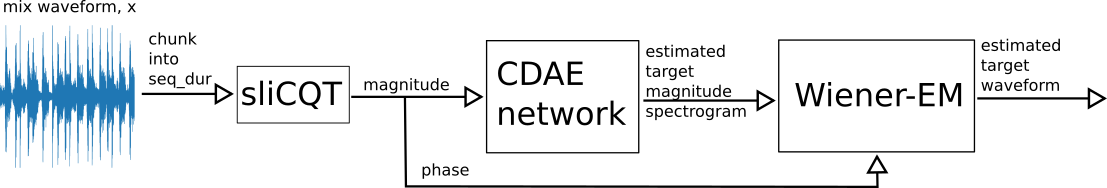}}
	\caption{Comparing UMX and xumx-sliCQ}
	\label{fig:blockdiagrams}
\end{figure*}

Our model, xumx-sliCQ, was trained on MUSDB18-HQ. On the test set, xumx-sliCQ achieved a median SDR of 3.6 dB versus the 4.64 dB of UMX and 5.54 dB of X-UMX, performing worse than the original STFT-based models. The overall system architecture of xumx-sliCQ is similar to UMX and X-UMX, shown in Figure \ref{fig:blockdiagrams}.

\section{Acknowledgements}
Thanks to my colleagues N{\'e}stor N{\'a}poles L{\'o}pez and Timothy Raja de Reuse, and to my master's thesis supervisor Prof. Ichiro Fujinaga, for help throughout the creation of xumx-sliCQ. Thanks to the MDX 21 challenge for creating a motivational environment for learning more about music demixing.
\clearpage\vfill
\flushend
\bibliographystyle{IEEEbib}
\bibliography{refs}

\end{document}